\documentclass[aps,prd,twocolumn,showpacs,superscriptaddress,letterpaper,amsmath,amssymb,nofootinbib]{revtex4-2}

\usepackage{graphicx}
\usepackage{dcolumn}

\usepackage{amssymb}
\usepackage{amsmath}
\usepackage{gensymb}
\usepackage{bbm}
\usepackage{bm}
\usepackage{svg}
\usepackage{siunitx}
\usepackage{rotating}

\usepackage[normalem]{ulem}

\usepackage{fancyhdr}
\usepackage{parskip}
\usepackage{dcolumn}
\newcommand{\galprop}{\mbox{GALPROP}}

\usepackage{mhchem}
\newcommand{\CTw}{$\ce{CO}$}
\newcommand{\CTh}{$\ce{^{13}CO}$}

\newcommand{\hi}{$\ce{H\textsc{i}}$}

\newcommand{\htwo}{$\mathrm{H_{2}}$}

\newcommand{\gray}{$\gamma$-ray}
\newcommand{\grays}{$\gamma$-rays}
\newcommand{\fermilat}{{\it Fermi}--LAT}

\begin{document}

\preprint{APS/PRD}

\title{Deep Learning Models of the Discrete Component of the Galactic Interstellar \gray~Emission}

\author{Alexander Shmakov}
\email{ashmakov@uci.edu}
\author{Mohammadamin Tavakoli}
\email{mohamadt@uci.edu}
\author{Pierre Baldi}
\email{pfbaldi@uci.edu}
\affiliation{Department of Computer Science, University of California, Irvine, California 92697, USA}
\author{Christopher M. Karwin}
\email{ckarwin@clemson.edu}
\affiliation{Department of Physics and Astronomy, Clemson University, Clemson, South Carolina 29634, USA}
\affiliation{Department of Physics and Astronomy, University of California, Irvine, California 92697, USA}
\author{Alex Broughton}
\email{abrought@uci.edu}
\affiliation{Department of Physics and Astronomy, University of California, Irvine, California 92697, USA}
\author{Simona Murgia}
\email{smurgia@uci.edu}
\affiliation{Department of Physics and Astronomy, University of California, Irvine, California 92697, USA}

\date{\today}

\begin{abstract}

A significant point-like component from the small scale (or discrete) structure in the  \htwo{} interstellar gas might be present in the \fermilat{} data, but modeling this emission relies on observations of rare gas tracers only available in limited regions of the sky. Identifying  this contribution is important to discriminate  \gray{} point sources from interstellar gas, and to better characterize extended \gray{} sources.  We design and train convolutional neural networks to predict this emission where  observations of these rare tracers do not exist, and discuss the impact of this component on the analysis of the \fermilat{} data. In particular, we evaluate prospects to exploit this methodology  in the characterization of the \fermilat{} Galactic center excess through accurate modeling of point-like structures in the data to help distinguish between a point-like or smooth nature for the excess. We show that deep learning may be effectively employed to model the  \gray{} emission traced by  these rare \htwo{}  proxies within statistical significance in data-rich regions, supporting prospects to employ these methods in yet unobserved regions.
 
\end{abstract}
\maketitle

The Galactic \gray~interstellar emission (IE) traces interactions of cosmic rays (CRs) with the interstellar gas and radiation field.  In a  companion paper~\cite{companion}, we showed that interstellar $\mathrm{H_2}$ gas is more structured and point-like than current IE models assume, and the related \gray~emission  might be a  statistically significant component  of the \fermilat{} data. If this structure is not adequately captured by the IE model,  it can  impact the  identification of resolved point-like sources as well as the characterization of  extended components   in the \gray~sky. We demonstrated that  unidentified  sources in the the fourth \fermilat{} catalog~\cite{2020arXiv200511208B}   could indeed be originating from it. In addition, we have argued that this  component   could artificially inflate the   unidentified and/or unresolved point source component  in the data and, depending on its morphology, contribute to confounding the interpretation of the  the Galactic center (GC) excess  observed by \fermilat~\cite{companion,goodenough2009,Hooper:2010mq,PhysRevD.84.123005,abazajian2011,abazajian2014,PhysRevD.88.083521,daylan2014characterization,calore2014background,ajello2016fermi} (see~\cite{Murgia:2020dzu} for a review). Improved modeling of small scale  gas related emission in the \fermilat{} data is therefore necessary to robustly characterize  \gray{} sources.

Interstellar \htwo{} gas  is traced indirectly via the   emission lines of other molecules that are  found concurrently in gas clouds. 
Carbon monoxide ($\mathrm{^{12}CO}$, or  \CTw ~hereafter)  is  used as a proxy but it is optically thick in  the  denser regions of molecular clouds, and therefore it  underestimates the total \htwo{} column density there. \CTw{} isotopologues, such as \CTh, although rarer, are not as optically thick and therefore more reliable to probe  dense \htwo~cloud cores, and therefore the  \htwo{} small scale structure. 
We briefly summarize the methodology we developed in ~\cite{companion} to model this emission. We employ observations of the  $J$ = 1--0 transitions of \CTw~and  \CTh~from the Mopra Southern Galactic Plane CO Survey~\cite{Braiding_2018}, which cover a 50 square degree region, spanning Galactic longitudes  $l$=300--350$^\circ$ and latitudes $|b|\le$ 0.5$^{\circ}$.  
 We calculate \htwo~column densities corresponding to Mopra's  \CTw{} and \CTh{}, referred to as    {$N(\mathrm{H_2})_{\mathrm{CO}}$} and {$ N(\mathrm{H_2})_{\mathrm{CO13}}$} respectively.  
  With this as input, we build a Modified  Map  by replacing   {$N(\mathrm{H_2})_{\mathrm{CO12}}$} with   {$N(\mathrm{H_2})_{\mathrm{CO13}}$} for pixels where    {$N(\mathrm{H_2})_{\mathrm{CO13}}$} $>$ {$N(\mathrm{H_2})_{\mathrm{CO12}}$}. The ``Modified Excess Template", which traces the small-scale $\mathrm{H_2}$-related \gray~ emission traced by \CTh{},  is determined from   the difference between the  Modified Map and the baseline \CTw~map.

 \begin{figure}[t] 
    \centering
    \includegraphics[width=0.495\textwidth]{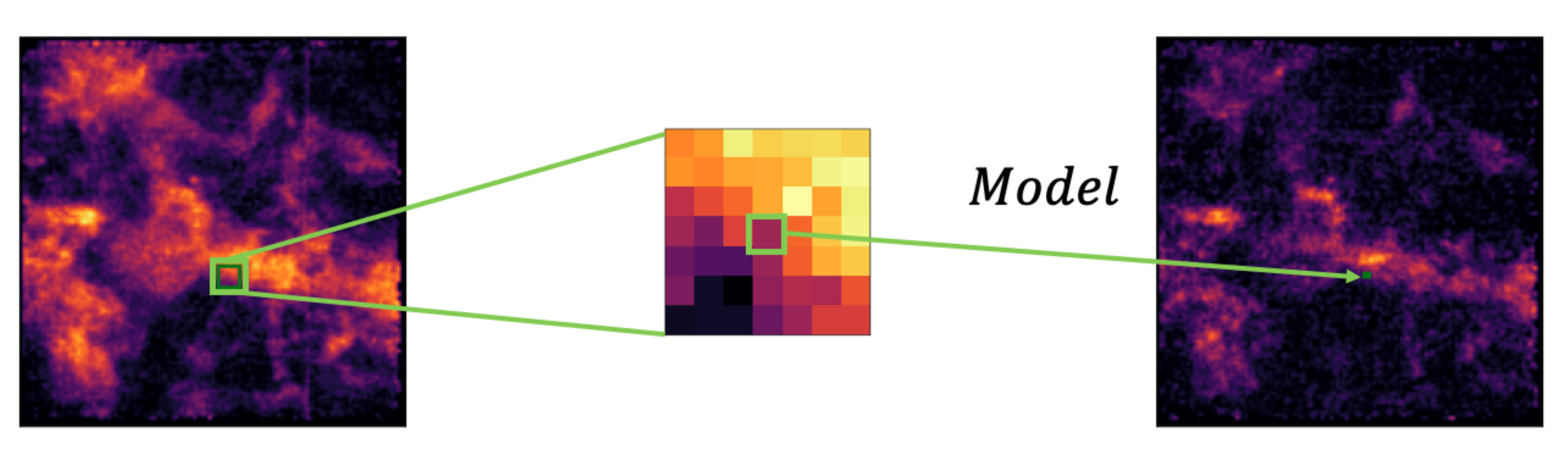}
    \caption{Graphical representation of the simplified modeling task demonstrating splitting a \CTw{} column density $\mathrm{1^{\circ}\times1^{\circ}}$ region ({\it left}) into smaller patches ({\it center}) which are associated with their respective target values in the   \CTh{} region ({\it right}).}
    \label{fig:patch}
\end{figure}
The central idea of this paper is to harness machine learning (ML), in particular deep learning \cite{baldi2021deep}, to predict the distribution of \CTh{} based on the \CTw{} observation, and therefore infer the \htwo~small scale structure in   regions  where \CTh~observations do not exist. Since a straightforward and robust analytical mapping between the two distributions is not available, ML could estimate this mapping from data. We train a deep learning model to map between \CTw{} and  \CTh{} column densities in the Mopra region. To simplify this regression problem and enlarge our effective data set, we subdivide the maps into small $D_1 \times D_2$ sections (patches) of the sky to their respective center points (Fig.~\ref{fig:patch}). The validity of this simplification requires the gas column density to be locally correlated. That is, we assume it is unlikely that pixels outside of our chosen patch  will significantly change the \CTh{} estimate. This simplification greatly reduces the model size and provides us with an effective data set of over $50$ million patches when $D_1 = D_2 = 0.0581 \degree$, or 7 pixels. We find that larger patch sizes do not improve model accuracy, justifying this assumption. We also apply smoothing techniques to eliminate noise from the data. The \CTw{}$\rightarrow$\CTh{} modeling problem may now be formally written as finding a parameterized function $f_\theta$ mapping a source region $S \in \mathbb{R}^{D_1 \times D_2}$ to an estimate target column density at the center of that same region $T \in \mathbb{R}$. The source $S_i$ corresponds to Mopra \CTw{} column densities and the targets $T_i$ to the corresponding \CTh{} column densities at the center of the patch (Fig.~\ref{fig:patch}). 
We evaluate our estimates by splitting the Mopra region into independent sub-regions which we designate for training and testing. We employ different splitting choices, but we report on one, the \textit{Alternate Tiled} split, throughout the remainder of this paper. This split consists of alternating $1^{\circ}\times1^{\circ}$ regions (or {\it tiles}), training on tiles containing longitudes $\{ 301-302, \dots, 349-350 \}$ and testing on tiles containing $\{ 300-301, \dots, 348-349 \}$. This allows us to evaluate the model's performance for neighbouring regions, reducing the variance between training and testing data distributions.

\begin{figure}[t] 
    \centering
    \includegraphics[width=0.49\textwidth]{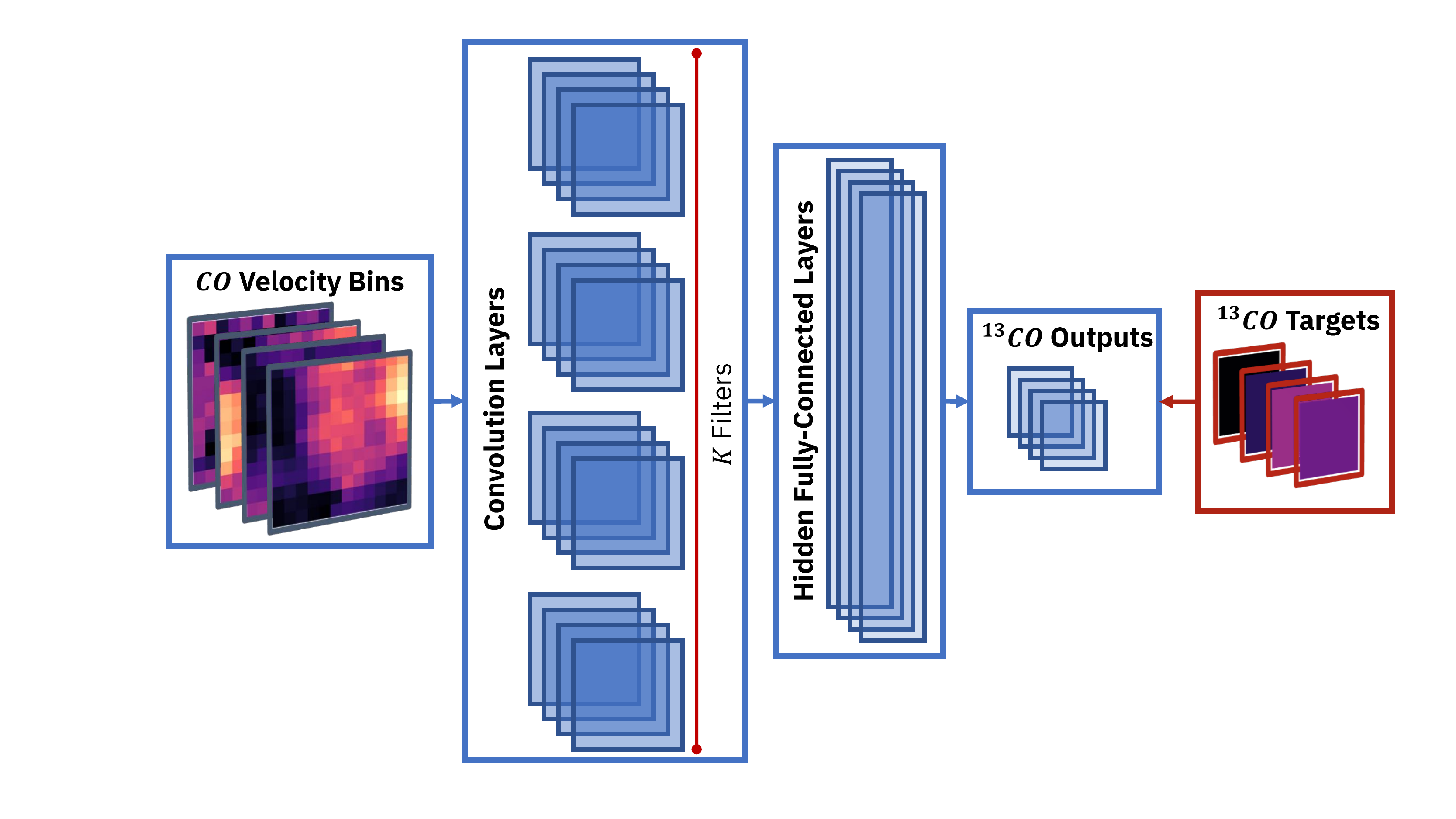}
    \label{fig:network}%
    \caption{A diagram of the internal layers of the network. For simplicity, only 4 velocity bins are shown, whereas the full network uses all 17 velocity bins. The fully connected layers share weights and the same layers are applied to every velocity bin.}
    \label{fig:network}
\end{figure}

\begin{figure*}[t]
\centering
\includegraphics[width=1.0\textwidth]{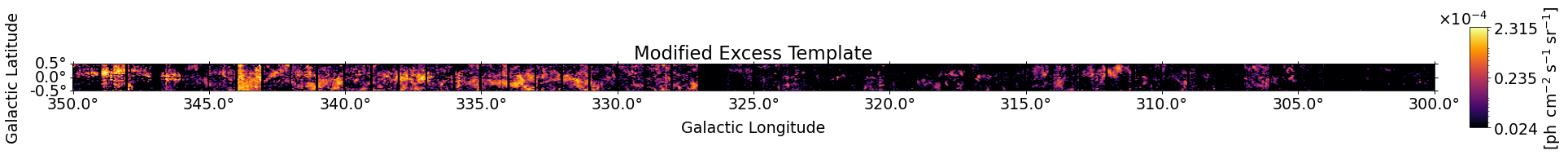}
\includegraphics[width=1.0\textwidth]{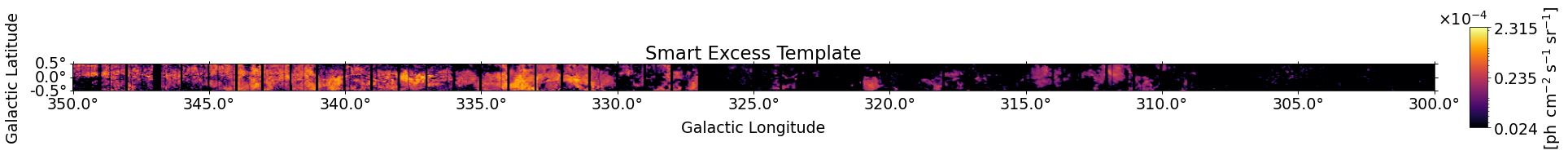}
\caption{  {\it Top}: the Modified Excess Template (from~\cite{companion}). {\it Bottom}: the Smart Excess Template. The color scale (logarithmic) indicates the \gray~intensity per  0.03125$^2$ degree pixel.  The first tile to the left (closest to the GC, and covering longitudes ${ 349-350}$) is testing data, and the neighboring one is training data for the CNN. The interleaving pattern of testing and training tiles is adopted for the entire Mopra region. 
}
\label{fig:smart_excess_template}
\end{figure*}
We employ convolutional neural networks (CNNs) to model and predict the \CTh{} column density in a translation-invariant manner. CNNs use neurons with shared connection parameters in order to implement convolution operations that provide the basis for building translation invariant architectures
\cite{baldi2021deep}. Each convolution operation is associated with a \textit{kernel}, or filter, corresponding to a set of connection weights that are shared by all the neurons in the corresponding layer.  
 For each of the $17$ velocity bins, we apply $K$ learnable convolution filters of size $D_1 \times D_2$, operating on an entire patch and independently for each velocity bin, producing $17$ $K$-dimensional latent vectors for each patch.  Afterwards, we apply parametric rectified linear units ($PReLU$) \cite{agostinelli2014learning, prelu, tavakoli2021splash} with a learnable slope, a batch normalization layer \cite{batchnorm}, and a random dropout layer \cite{srivastava2014dropout, baldidropout14}. The latent vectors are then processed by several fully-connected layers, each with their own $PReLU$, batch-normalization, and dropout. Unlike the convolution layer, these hidden layers are shared between velocity bins, learning identical weights for every bin. This design allows spatial components (CNN) of the network to be specialized for each velocity bin while allowing latent higher level components to be shared between bins. The resulting latent vectors are fed through a final fully-connected layer which produces the \CTh{} concentration point estimates for each patch and velocity bin. A diagram of the network architecture is presented in Figure \ref{fig:network}.

\begin{figure*}
\centering
\includegraphics[width=1.006\textwidth, scale=1.2]{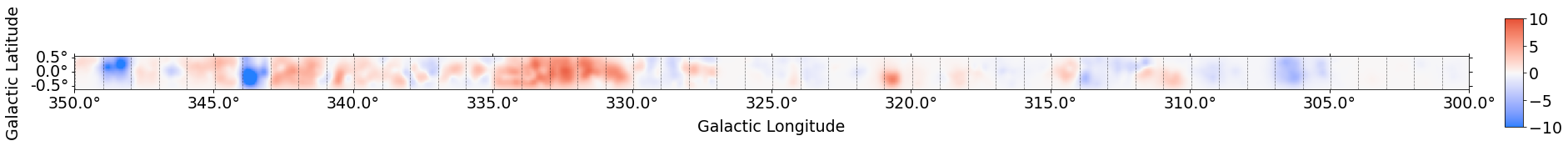}
\includegraphics[width=1.00\textwidth, scale=1.2]{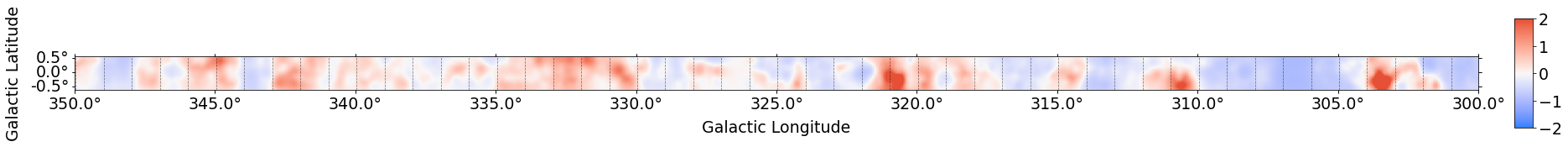}
\includegraphics[width=1\textwidth, scale=1.2]{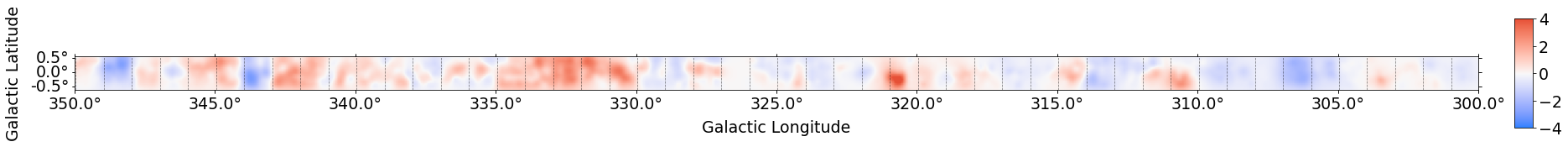}
\caption{The pixel-wise residuals for the full region between the predicted counts from the Smart Excess Template and the predicted counts from the Modified Excess Template. \textit{Top}: residual in counts (Smart-Modified), \textit{Middle}: fractional residuals ((Smart-Modified)/Modified), \textit{Bottom}: Residual in units of  1 standard deviation for the Modified Excess Template ((Smart-Modified)/$\sqrt{\mathrm{Modified}}$). The color scale is per  0.08$^2$ degree pixel.}
\label{fig:ExcessTemplate_residuals}
\end{figure*}

The Mopra dataset contains varying regions of column densities from bright to very dim. To effectively learn this high-spread distribution, we model the incidence of photons on the Mopra detector as a Poisson process and use a Poisson log-likelihood loss. We further re-weight this loss based on the \CTh{} density to elevate the importance of bright pixels, prioritizing accuracy in hot-spots over a slight degradation in the background. This increases the importance of accurate measurement in the bright regions, encouraging the network to focus on the accuracy of these regions first. As the network trains, we anneal this weighting back to uniform (with all target values having equal weighting) in order to minimize any bias introduced by this loss. The design of this loss is guided by the overall goal of finding small angular scale features while limiting the amount of overestimation throughout the Mopra region. 

We tune the CNN's hyperparameters using the SHERPA hyperparameter optimization library \cite{sherpa,hertel2020sherpa}, testing 2000 network variations, scoring each parameterization with Poisson likelihood, and using Gaussian Process optimization to suggest parameters. Our final parameterization is evaluated after training the network for $200$ iterations.
We generate the predicted \CTh{} density by stitching the network output for every patch in a source image. When performed on a single Nvidia Titan XP GPU, and batching $8192$ patches at once, this inference process requires approximately $11$ seconds to cover the entire Mopra region.

\begin{figure*}
\centering
\includegraphics[scale=0.24]{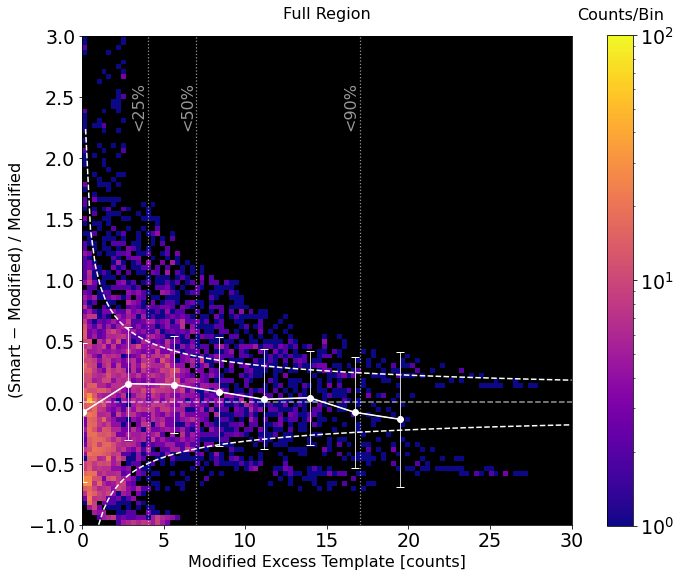}
\includegraphics[scale=0.24]{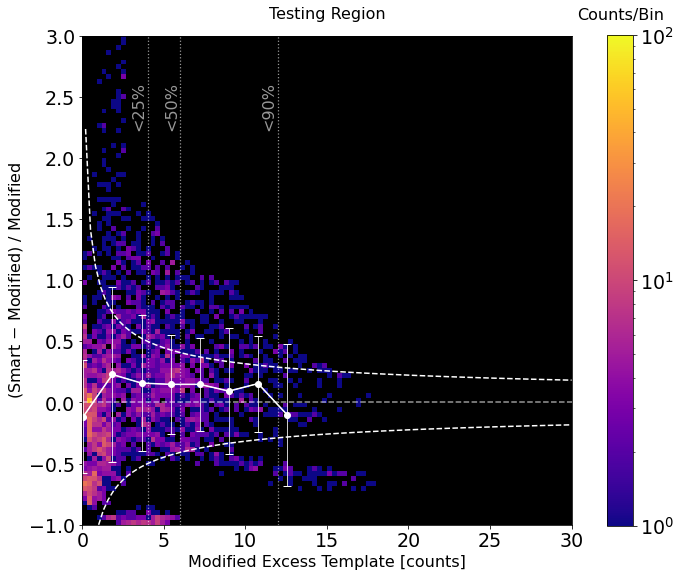}
\includegraphics[scale=0.24]{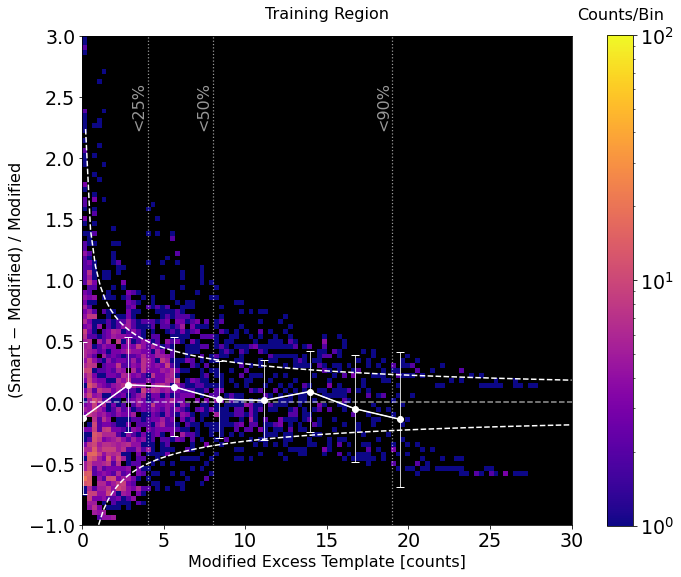}
\caption{The fractional residuals between the nominal Smart Excess Template (\textit{Smart}) and Modified Excess Template (\textit{Modified}) predicted counts per pixel  as a function of Modified Excess Template pixel counts for all tiles ({\it left}), testing tiles ({\it center}), and training tiles ({\it right}).  The points and error bars represent the median and median absolute deviation of the fractional difference for the pixels in 8 domain intervals (we note that the scatter of the points within each interval is not simply characterized by these quantities and does not always indicate a most probable outcome.) We also plot the boundaries of the $\pm1\sigma$ of the Poisson noise fluctuations of the Modified Excess Template. We  overlay the 25\%, 50\%, 95\% flux percentiles to indicate the flux fraction encompassed by pixels below that threshold.}
\label{fig:BinnedResiduals_ExcessTemplate_residual}
\end{figure*}

Following the same procedure as in~\cite{companion}, we employ the CR propagation code \galprop{} (v56)\cite{Moskalenko:1997gh,Moskalenko:1998gw,Strong:1998pw,Strong:1998fr,2006ApJ...642..902P,Strong:2007nh,Vladimirov:2010aq,Johannesson:2016rlh,porter2017high,Johannesson:2018bit,PhysRevC.98.034611} to calculate $\gamma$-ray sky maps in 17 Galactocentric radial bins for the $\mathrm{H_2}$-related emission, where the latter is  traced by \CTw~and \CTh. 
The   \CTw~ and \CTh~ related \htwo~column densities, {$ N(\mathrm{H_2})_{\mathrm{CO13}}$} and {$N(\mathrm{H_2})_{\mathrm{CO}}$} respectively, are used to determine    the \htwo~that is missed in dense regions when only CO is used as a tracer.  
 We follow  the  procedure from~\cite{companion}  to determine the Modified Map and  ``Modified Excess Template" (Fig.~\ref{fig:smart_excess_template}, top panel)   to construct their ML analogs. The ``Smart Map" is determined  similarly to the Modified Map, but instead of using the true Mopra \CTh~ data, we use the \CTh~ estimates from the CNN. We use this to determine the  "Smart Excess Template", defined as the difference between the Smart map and the baseline \CTw~map.  The Smart Excess Template, also integrated over all annuli and energies, is shown in Fig.~\ref{fig:smart_excess_template}, bottom panel. The map covers the full Mopra region, which includes alternating  training and testing tiles for the CNN. 
  
To assess how closely the CNN predicts the \gray~ emission inferred by Mopra observations, we compare the excess templates with three different metrics. We quantify the differences with the following three metrics which are determined by using the predicted photon counts per pixel for each of the templates and scaled to $\sim$12~years of \fermilat~data (we assume the same \fermilat~observation parameters and event selection as for the simulations described in ~\cite{companion} and in this work): absolute difference (Smart-Modified), fractional difference ((Smart-Modified)/Modified), difference in units of the standard deviation, $\sigma$, for the Modified counts ((Smart-Modified)/$\sqrt{{\mathrm {Modified}}}$). The latter metric allows us to compare the magnitude of the differences due to the CNN performance in light of the statistical power of the data. All results are shown in Fig.~\ref{fig:ExcessTemplate_residuals} as a function of longitude and latitude for the  Mopra region. We observe that for the vast majority of the  region (83.6\% of the pixels), the predicted counts for the Smart Excess Template are within $\pm1\sigma$ of the Modified Excess Template counts, and therefore the difference is generally  within  the statistical uncertainty of the data. 
This result indicates that  the CNN performance is adequate for modeling the small scale structure in \htwo-related \gray~emission traced by Mopra, for the statistics achieved by \fermilat.

The  CNN performance varies with respect to longitude. Most tiles display either an overall underprediction by the CNN across the entire tile or an overprediction, rather than a comparable mixture of under/over predictions across the same tile. This may be explained by the design of the loss function, placing increased importance towards brighter pixels while also biasing the network to prefer underprediction via Poisson regression. Prominent examples include the training tile $343^{\circ} < l < 344^{\circ}$ and testing tile $348^{\circ} < l < 349^{\circ}$, the brightest  in each dataset, where the CNN underpredicts the gas column density, as shown in Fig.~\ref{fig:ExcessTemplate_residuals}. 
The dependence of the residuals as a function of pixel brightness (in counts/pixel), is shown in Fig.~\ref{fig:BinnedResiduals_ExcessTemplate_residual} for the full Mopra region, and for the training and testing regions separately. Generally, better agreement between Mopra (Modified Excess Template) and the  CNN prediction (Smart Excess Template) is found for the brighter pixels. There is a broad spread in the fractional residuals, but it  is confined to be between $\pm$50\% for the vast majority of the pixels.  We overlay contours to indicate the $\pm$1$\sigma$ statistical fluctuations in the Modified Excess Template, which encompasses 83.6\% of the pixels.  We also  overlay the median and median absolute deviation  of the fractional difference in bins of the Modified Excess Template counts to guide the eye, but emphasize  that the scatter of the points within each  interval cannot be simply characterized by these quantities and does not always indicate a most probable outcome. As expected, we find that the CNN performs (somewhat) better in the training tiles. 
Approximately 58\% of pixels where the CNN prediction is beyond  the $\pm1\sigma$ level  are in the testing regions. The CNN is more likely to overpredict the emission for the dimmer pixels (60.8\%  of pixels across the full region are below the 25\% flux percentile), consistently with the design. The CNN overprediction in the dim pixels   essentially spreads  out the brightness of hot-spots over a larger area.   We also note that small statistics causes the fractional difference to increase dramatically for the faintest pixels.  Above the 90\%  flux percentile, the  distribution of the fractional residuals bifurcates into two separate distribution. The underpredicted pixels are in  the $343^{\circ} < l < 344^{\circ}$ tile, and the overpredicted pixels are in  $333^{\circ} < l < 334^{\circ}$ tile, both in the training regions. This is likely caused by our choice to combine independent convolution layers for each velocity bin with a shared hidden layer. Since certain bins have higher overall brightness than others, the shared hidden layer will attempt to average the error between the two extremes. 

We determine the significance of the Smart Excess Template  in the \fermilat~data, similarly to~\cite{companion} for  the Modified Excess Template. The simulations cover the same observations and event selection ($\sim$12 years, $1-100$~GeV, P8R3 CLEAN FRONT+BACK).  We  only simulate the $\mathrm{H_2}$-related \gray~emission in the Mopra region,  excluding all other components, since our goal is to  establish  the performance of the Smart Excess Template in the optimistic scenario where  all other components are known. The  simulated events  trace the $\mathrm{H_2}$-related $\gamma$-ray emission modeled with the Modified Map from Mopra. The simulated data are  fit based on a binned maximum likelihood method to a model that  includes  the  baseline \CTw~map  (as observed by Mopra) and  the Smart Excess Template. The normalization of the Smart Excess Template is free to vary in the fit. The energy spectrum, calculated by GALPROP, is  assumed, and held fixed during the likelihood fit.  The normalization and spectral index of the \CTw~baseline contribution is also free to vary. The 17 radial bins from GALPROP are combined into 4 radial annuli, which  we refer to as A1, A2, A3, and A4, with the same partitioning as in~\cite{companion}. Also consistently to~\cite{companion}, the normalization of A4 is held constant to a normalization of 1.0, contributing a partial flux of $4.98\times10^{-9}$ \si{ph.cm^{-2}.s^{-1}}.

We simulate 1000 realizations  and calculate the Test Statistics (TS) for the nested models ($\mathrm{-2log(L_0/L)}$, where $\mathrm{L_0}$ is the null hypothesis (CO baseline), and $\mathrm{L}$ is the alternative hypothesis (CO baseline and Smart Excess Template.) The statistical significance is approximated by $\sigma \approx \sqrt{\mathrm{TS}}$.  
\begin{figure}
\centering
\includegraphics[width=0.5\textwidth]{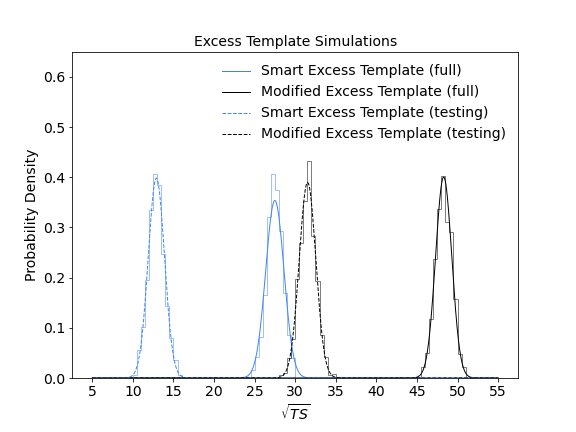}
\caption{Statistical significance of the Smart Excess Template for the full Mopra region (solid lines),  and for the testing regions (dashed lines). Each contains 1000 realizations of 12 years of \fermilat~data. A fit with a Gaussian distribution is overlaid to each distribution. These distributions are compared to the results for the Modified Excess Template from~\cite{companion}.}
\label{fig:excess_significance}
\end{figure}
\begin{figure*}
\centering
\includegraphics[width=1.0\textwidth]{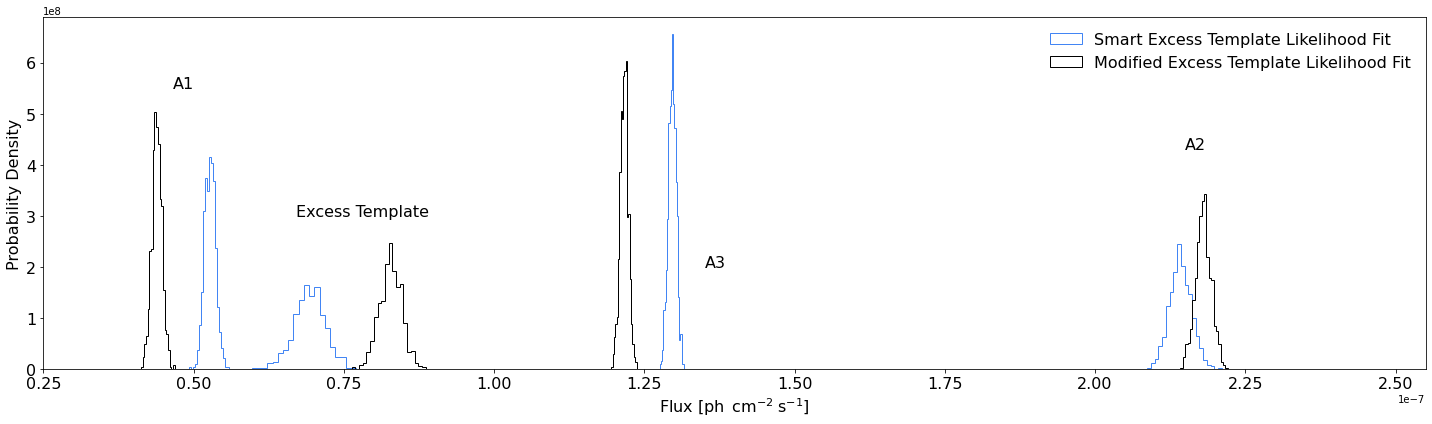}
\caption{Distribution of fluxes for each  component of the best-fit model using the Smart Excess Template maps over the full region for 1000 simulations of \fermilat~data. The results for the Modified Excess Template maps from~\cite{companion} are overlaid.  Note that one of the annuli (A4), not shown,   has negligible contribution and the normalization was fixed in the fitting process.}
\label{fig:main_hist2}
\end{figure*}
The distribution of the $\sqrt{TS}$ for the  simulations is shown in  Fig.~\ref{fig:excess_significance}  for the full Mopra region (solid line), and for the testing regions only (dashed line). 
The Smart Excess Template corresponds to $\sqrt{TS}=27.5 \pm 1.1$ (mean and standard deviation) in the full Mopra region and $12.9 \pm 1.0$ in the testing region. The significance in the testing region is lower, not only because of the smaller statistics, but  also because  the full region  contains the training tiles  where the CNN more closely matches the Mopra data. For comparison,  the Modified Excess Template has a  $\sqrt{TS}=48.3 \pm 1.0$ and $\sqrt{TS}=31.5 \pm 1.0$  for  the full  and  testing regions, respectively.  
\begin{figure}
\centering
\includegraphics[width=0.5\textwidth]{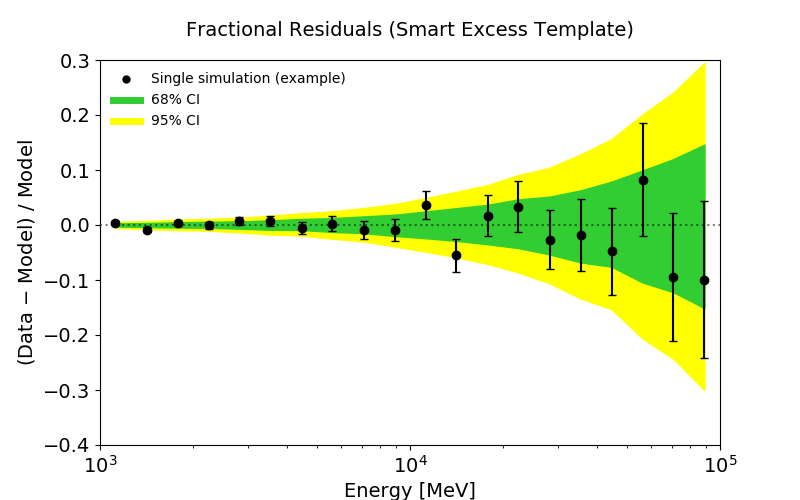}
\caption{Fractional count residuals for the fit with the Smart Excess Template. The green and yellow bands show the 68\% and 95\% confidence regions from 1000 simulations, respectively. We also plot the results for a single simulation as an example in each case, which is shown with black data points.}
\label{fig:gof}
\end{figure}
\begin{figure*}
\centering
\includegraphics[width=1.009\textwidth, scale=1.2]{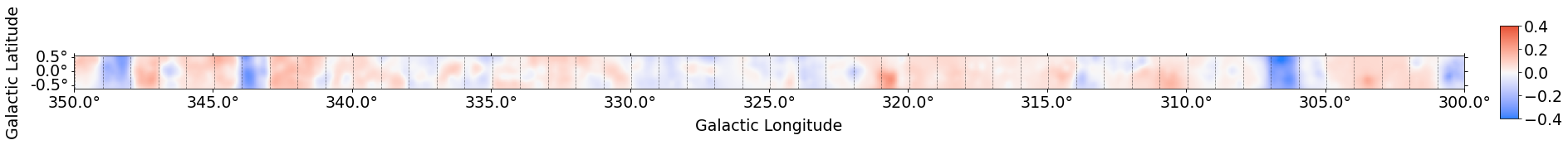}
\caption{Fractional count residuals, (Data-Model)/Model,  in latitude and longitude for the fits with the Smart Excess Template, in addition to the other \CTw{} components. The residuals are calculated using  the mean over 1000 simulations.}
\label{fig:residuals}
\end{figure*}
In Fig.~\ref{fig:main_hist2}, we show the distributions for the best-fit  flux 
of the Smart Excess Template overlaid with those for the Modified Excess Template, as well as the flux for the CO baseline emission, separated into annuli.  The integrated flux for the Smart Excess Template is approximately 83.7\% of the flux of the Modified Excess Template, indicating an overall underprediction  by the CNN.  
The fractional count residuals ((Data-Model)/Model) as a function of energy for the best fit model are shown in Fig.~\ref{fig:gof} for the fits that include Smart Excess Template, in addition to the other \CTw{} components. They are consistent with zero. The best fit spectra of the \CTw{} components agree with the GALPROP prediction.  Finally, Fig.~\ref{fig:residuals}  shows the fractional count residuals in latitude, longitude. The residuals, which incorporate differences between the (best fit) Smart Excess Template and the (simulated) Modified Excess Template,  are smaller than those shown in Fig.~\ref{fig:ExcessTemplate_residuals}, (middle panel) indicating the other CO model  components partially compensate for the discrepancies between the Excess Templates. We have performed this analysis also  using Gaussian Processes and find that the performance is worse compared to the CNN.

The ML methodology presented here must be refined to be extended to other regions of the sky. The available all sky  \CTw{} $J$=1--0 observations have significantly  poorer  spatial resolution (8$^{\prime}$ from~\cite{2001ApJ...547..792D}) compared to Mopra ($0.6^{\prime}$). The CNN model will therefore require additional transfer learning to operate on lower resolution maps. The resolution of the \fermilat~data is worse (for most energies and event types), suggesting that the impact of the poorer \CTw{} resolution on the CNN predictions for the \grays{}  might ultimately be less pronounced. Another consideration is that the available \CTh{} training data for the CNN (from  Mopra, and other  observations~\cite{2018MNRAS.473.1059U,2020MNRAS.498.5936E,2020A&A...639A..26K,2017IAUS..322..164B,2018ApJ...855...33D} that could also be included), are  confined to the Galactic plane. The \htwo{} scale height is low, but its small scale  structure contribution at higher latitudes might not be negligible because of the contribution of more local \htwo{}. However,  because of the lack of adequate training data the CNN prediction for this component could be more uncertain and  this  is relevant for the characterization of the GC excess, which extends to high latitudes.  Finally, this analysis inherits the limitations and uncertainties in modeling the \htwo{} component with traditional methods, including the loss of kinematic resolution for the gas in the direction of the GC~\cite{ackermann2012fermi}. And it adds another: the $\mathrm{H_2}$ column densities from  \CTw~and \CTh{} have been treated independently in this work.  Their estimates can be related analytically, but  rely on  a  assumptions pertaining to the optical depth, beam filling factor, spatial variation, etc.~(e.g.~\cite{burton2013mopra,10.1093/mnras/sty059}). 
More ambitiously,  ML may be used to constrain some of these uncertainties by using  the \gray~data. Finally,  small scale structure in the \gray{} data could also arise from other components of the IE, e.g. related to \hi{}, and shall be included in a more comprehensive study.

\textit{Conclusions}.--- We present a methodology that harnesses ML to predict  the small scale  component of the interstellar \htwo{}  gas and its related \gray{} emission for the first time. Incorporating this contribution in \gray~IE models   is  crucial  as it impacts the determination of \gray{} point sources and could affect characterization of  extended \gray{} emissions, e.g. the \fermilat{} GC excess.  To this end, we employ observations of tracers of this emission, which are only available in limited regions of the sky (covered by the Mopra survey), and train a  CNN to predict this component elsewhere.
We have tested the performance of this methodology  in predicting the contribution of the \htwo{} related small scale structure  to the \fermilat{} data and  conclude that deep learning can  model the  \gray{} emission well in data-rich regions supporting prospects to employ and extend  this methodology to other regions of the sky.

\section{Acknowledgements}
We thank Troy Porter for many helpful discussions and insights. The work of AS, MT, and PB is in part supported by grants NSF NRT 1633631 and ARO  76649-CS to PB.

\bibliography{Bibliography}

\end{document}